\documentclass[showpacs,twocolumn,aps,prb,superscriptaddress]{revtex4-1}
\usepackage[colorlinks, hyperfigures, linkcolor=black, citecolor=black, breaklinks,]{hyperref}
\usepackage[centertags]{amsmath}
\usepackage{amssymb,eqnarray}
\usepackage{amsthm}
\usepackage{fancyhdr}
\usepackage{graphicx,subfigure}
\usepackage{epsfig}
\usepackage{epsfig}
\usepackage{amssymb}
\usepackage{bbm}
\usepackage{graphicx}
\usepackage{amsmath}
\usepackage{subfigure}
\usepackage{bm}

\newcommand{\om}{\Omega}
\newcommand{\omdag}{{\Omega}^{\dagger}}
\newcommand{\omeq}{\Omega_{\rm eq}}

\newcommand{\delom}{\delta\Omega}

\newcommand{\F}{\boldsymbol{F}}

\newcommand{\pp}{\boldsymbol{p}}

\newcommand{\p}{\boldsymbol{p}}

\newcommand{\rr}{\boldsymbol{r}}
\newcommand{\nab}{\boldsymbol{\nabla}}
\newcommand{\xxi}{\boldsymbol{\xi}}

\newcommand{\eeta}{\boldsymbol{\eta}}
\newcommand{\R}{\boldsymbol{R}}

\begin{document}
		\title{Linear response approach to active Brownian particles
                  in time-varying activity fields}
\author{Holger Merlitz}
\email{merlitz@posteo.de}
\affiliation{Leibniz-Institut f\"{u}r 
Polymerforschung Dresden, Institut Theorie der Polymere, 01069 Dresden}
\author{Hidde Vuijk}
\affiliation{Leibniz-Institut f\"{u}r 
Polymerforschung Dresden, Institut Theorie der Polymere, 01069 Dresden}
\author{Joseph Brader}
\affiliation{Department of Physics, University of Fribourg, CH-1700 Fribourg,
  Switzerland}
\author{Abhinav Sharma}
\affiliation{Leibniz-Institut f\"{u}r 
Polymerforschung Dresden, Institut Theorie der Polymere, 01069 Dresden}
\author{Jens-Uwe Sommer}
\affiliation{Leibniz-Institut f\"{u}r 
Polymerforschung Dresden, Institut Theorie der Polymere, 01069 Dresden}
\affiliation{Technische Universit\"{a}t Dresden, Institute of Theoretical Physics, 01069 Dresden, Germany}

%\textsuperscript{1}, Hidde Vuijk\textsuperscript{1}, Joseph Brader\textsuperscript{2}, Abhinav Sharma\textsuperscript{1}, Jens-Uwe Sommer\textsuperscript{1,3}}
%		{\footnote{email:merlitz@posteo.de}}
%	\affiliation{\textsuperscript{1} Leibniz-Institut f\"{u}r Polymerforschung Dresden,
%		Institut Theorie der Polymere, 01069 Dresden\\
%		\textsuperscript{2} Department of Physics, University of Fribourg, CH-1700 Fribourg, Switzerland\\
%		\textsuperscript{3} Technische Universit\"{a}t Dresden, Institute of Theoretical Physics, 01069 Dresden, Germany 
%		}

\begin{abstract}
In a theoretical and simulation study, active Brownian particles (ABPs)
in three-dimensional bulk systems are exposed to time-varying 
sinusoidal activity waves that are running through the system. A 
linear response (Green-Kubo) formalism is applied to derive fully
analytical expressions for the torque-free polarization profiles
of the particles. The activity
waves induce fluxes that strongly depend on the particle size and
may be employed to de-mix mixtures of ABPs or to drive the particles 
into selected areas of the system. Three-dimensional Langevin dynamics 
simulations are carried out to verify the accuracy of the linear response 
formalism, which is shown to work best when the particles are small 
(i.e.\ highly Brownian) or operating at low activity levels.        
\end{abstract}
\maketitle

\section{Introduction}

Biological systems often contain microscopic objects of different
sizes, featuring a self-propulsion to play a crucial role in 
vital actions. Prominent examples are white blood cells chasing 
intruders~\cite{fenteany_COH04}, motor proteins facilitating 
the transport of e.g.\ RNA inside cells~\cite{kanai_Neuron04},
or bacteria such as Escherichia coli which are simply searching
for food~\cite{berg_04}.

On the other hand, synthetic microswimmers are designed
to offer hope-bearing solutions to problems that are hard
to approach with the toolbox available to conventional 
nanotechnology~\cite{ebbens_SM10,yamamoto_powder15}. 
Potential fields of applications are highly directional 
drug delivery, improved biomarkers or contrast 
agents~\cite{abdelmohsen_JMCB14}, as well as the elimination
of pollutants in the framework of environmental 
protection~\cite{gao_ACSNano14}.

By far the simplest synthetic self-propelling agents are
binary Janus-particles of spherical shape. These particles
may be driven by catalytic reactions with an ingredient of the
solvent such as hydrogen peroxide~\cite{howse_PRL07} or 
hydrazine~\cite{gao_JACS14}. If the 'fuel' exhibits a
concentration gradient, then the activity of these ABPs 
turns position dependent, leading to phenomena that have
recently been demonstrated to resemble 
chemotaxis~\cite{peng_AC15,ghosh_PRE15,vuijk_18}.
Instead of being chemically driven, Janus particles may
also respond to light as a result of an inhomogeneous
surface-heating and the resulting local temperature
gradient~\cite{lozano_NatureComm16}. This resembles
phototactic behavior, and since light intensities are
easily controlled in a laboratory setup, not only 
position-, but also rapidly varying time-dependent 
activity profiles may be employed to manipulate ABPs.
Recent reviews about the state of the art of artificial 
nano motors have been presented by Yamamoto et 
al.~\cite{yamamoto_powder15} and Bechinger et 
al.~\cite{bechinger_RevModPhys16}. 

Some of the authors have recently applied the linear response
(Green-Kubo) approach to ABPs and successfully computed average
swim speeds~\cite{sharma_JCP16} in homogeneous systems as well 
as a torque-free polarization and the resulting density 
distributions in systems with spatially  
inhomogeneous activity profiles~\cite{sharma_PRE17}. In the 
present work, we generalize that formalism to systems in
which the activity fields are additionally time dependent.
Recently, Geiseler et al.\ have analyzed the behavior of
microswimmers in active density waves that are sweeping
over the ABPs and thereby driving them into selected
directions~\cite{geiseler_PRE16,geiseler_SR17}. Techniques
to manipulate the global direction of motion of
ABPs are naturally of paramount interest to the practitioner.
While existing studies are restricted to two-dimensional
setups and thus affected by boundary effects of the substrate
on which the particles are sliding, our linear response 
approach is going to be applied to three dimensional bulk 
systems in absence of boundaries. 
  
In Sec.\ \ref{modelandtheory}, we first present our model,
the system of units and a short introduction to linear response
theory. Section \ref{sec:orientation} derives the orientation
profiles of ABPs that are exposed to a propagating sinusoidal 
activity wave. The resulting torque-free polarization profile 
inside the activity gradient is derived as a fully analytical
expression and compared to Langevin dynamics simulations. 
The corresponding density distributions are derived in Sec.\ \ref{sec:densities} 
and again a close agreement with the simulations is reported. 
Section \ref{sec:flux} studies the fluxes induced into the ABPs 
by the propagating activity wave, which strongly depend on 
the particle size. This fact is exploited in Sec.\ \ref{sec:separation} 
in which a mixture of ABPs of different sizes is de-mixed by
selective applications of induced fluxes. We discuss the
parameter ranges in which the linear response ansatz yields
accurate predictions (Sec.\ \ref{sec:activity}) and summarize
our findings in \ref{sec:summary}.

\section{Model and Theory}\label{modelandtheory}	

\subsection{Numerical simulations}

In this work, we introduce a standard particle with a diameter of $b = 1$
and mass $m = 1$, which define the length and mass units. We explicitly 
specify the translational diffusion coefficient $D_t = k_B T/\zeta$   
and the drag coefficient $\zeta$, and with $k_B = 1$, the unit-less temperature
of the system is specified, too. The unit time $\tau$ is the time which 
the standard particle needs to diffuse over the distance of its 
diameter, i.e.\ $\tau = (6D_t)^{-1}$. We have thus scaled the remaining
parameters to yield $\tau = 1$. 

The Langevin dynamics simulations were carried out with the 
molecular dynamics simulation
package LAMMPS~\cite{plimpton_CompPhys95} in fully three dimensional
setups. The activity is induced by an explicit force that acts in the 
direction of the particle's internal orientation vector, noise is
induced by a Langevin thermostat at a system temperature of $T = 5/3$. 
We use an integration time-step of $4\cdot 10^{-4}$. In the 
simulations of sections \ref{sec:orientation} -- \ref{sec:flux}, 
500 active particles are
placed into a box of the size $L^3 = 15^3$  and
periodic boundaries. The particles have a diameter of $b=1$,
but no interaction, so that actually an ensemble of single
particles is simulated. In Sec.\ \ref{sec:separation},
particles of different diameters are distributed inside a
larger box of size $15\times 15\times 150$, being bounded in
z-direction with impenetrable walls, but periodic in x- and
y-directions.
The remaining parameters are summarized in Tab.\ \ref{tab:parameters}.
Further technical details regarding the implementation of ABPs 
in LAMMPS have been described in our previous publication~\cite{merlitz_SM17}.

\begin{table}[htb]
%\renewcommand{\arraystretch}{1.1}%
%\begin{tabular}[b]{@{}r@{}l|l@{}}
\begin{tabular}[b]{c | c | c | c }
%\multicolumn{2}{@{}c|}{Quantity} & Meaning \\
diameter, $b $                  &  0.5      & 1    &     2 \\
\hline 
mass, $m $                  &  1/8      & 1    &     8 \\
frictional drag coefficient, $\zeta $      &  5        &  10   &     20 \\
momentum relaxation time, $\tau_m$         &  1/40     &  1/10 &     2/5 \\
translational diffusion coefficient, $D_t $     &  1/3     & 1/6 &   1/12 \\
rotational diffusion coefficient, $D_r$         &  4   &  1/2   &  1/16  \\
rotational relaxation time, $\tau_r $    &  1/8      &  1   &    8  \\
ratio, $\tau_r / \tau_m$                 &   5       &  10  &   20   \\
\hline
\end{tabular}
\caption{Simulation parameters for ABPs of different diameter $b$. The 
mass $m$ is required for the integrator, but the resulting momentum
relaxation time $\tau_m$ is significantly shorter than $\tau_r$ and therefore, 
the motion of the particle is overdamped on the time scale of $\tau_r$.  
\label{tab:parameters}}
\end{table}

With the choice of parameters as above, the motion of an active particle is
overdamped on the time scale of $\tau_r$. In the overdamped limit, the motion
of an active particle can be modelled by the Langevin equations
\begin{align}\label{full_langevin}
&\!\!\!\!\!\!\dot{\rr} = \frac{f(\rr, t)}{\zeta}\,\p  + \xxi\;\;,
\;\;\;
\dot{\p} = \eeta\times\p \,.
\end{align}
with coordinates $\rr$ and the embedded unit vector $\p$
which defines the particle orientation. $f$ is the modulus
of the force that drives the particle into the direction 
of its orientation vector, and
$\zeta$ is the frictional drag coefficient.
The stochastic vectors $\xxi(t)$ and $\eeta(t)$ are Gaussian distributed with zero mean and 
time correlations	
$\langle\xxi(t)\xxi(t')\rangle=2D_t\delta(t-t')$ and 
$\langle\eeta(t)\eeta(t')\rangle=2D_r\delta(t-t')$,
with the translational and rotational diffusion coefficients
$D_t$ and $D_r$, respectively. The latter is related to the
rotational relaxation time according to $\tau_r = 1/(2D_r)$. 
Note that in the second part of Eq.\ \eqref{full_langevin}, the
orientation vector does not couple to the activity field and hence
no direct torque acts on the particle due to the 
position-dependent activity. Our linear response, presented in the following subsection, is based on the overdamped set of equations~\eqref{full_langevin}. 

%
%We consider a three dimensional system of active, noninteracting, 
%spherical Brownian particles with coordinates $\rr$ and orientation 
%specified by an embedded unit vector $\p$. 
%A space-dependent self-propulsion force $f(\rr, t)$ 
%acts in the direction of orientation. 
%Omitting hydrodynamic interactions the motion can be 
%modelled by the Langevin equations
%%
%\begin{align}\label{full_langevin}
%&\!\!\!\!\!\!\dot{\rr} = \frac{f(\rr, t)}{\zeta}\,\p  + \xxi\;\;,
%\;\;\;
%\dot{\p} = \eeta\times\p \,.
%\end{align}
%where $\zeta$ is the frictional drag coefficient.
%The stochastic vectors $\xxi(t)$ and $\eeta(t)$ are Gaussian distributed with zero mean and 
%have time correlations	
%$\langle\xxi(t)\xxi(t')\rangle=2D_t\delta(t-t')$ and 
%$\langle\eeta(t)\eeta(t')\rangle=2D_r\delta(t-t')$
%with the translational and rotational diffusion coefficients
%$D_t$ and $D_r$, respectively. The latter is related to the
%rotational relaxation time according to $\tau_r = 1/(2D_r)$. 
%Note that in the second part of Eq.\ \ref{full_langevin}, the
%orientation vector does not couple to the activity field and hence
%no direct torque acts on the particle due to the 
%position-dependent activity.
%%

\subsection{Linear response ansatz}

It follows exactly from~\eqref{full_langevin} that the joint $N$-particle probability distribution $P(t) \equiv P(\bm{r}^N, \bm{p}^N, t)$ evolves according to~\cite{gardiner85} 
\begin{equation}
\frac{\partial P(t)}{\partial t} = \Omega_a(t) P(t)
\label{fp}
\end{equation}
with the time-evolution operator $\Omega_a$. 
%This operator 
%is then formally split into an equilibrium part, which
%accounts for the passive (diffusive) evolution of the system,
%and an active contribution which contains the self-propulsion
%of the particle. 
The time-evolution operator can be split into a sum of two terms, 
$\om_{\rm a}(t)=\omeq+\delom_{\rm a}(t)$, 
where the equilibrium contribution is given by 
\begin{align}\label{smol_op_eq}
\omeq = \sum_{i=1}^{N} \nab_{i}\!\cdot\!
\big[
D_\text{t}\!\left(\nab_{i}\! - \beta\F_i\right) 
\big] \!+\! D_\text{r}\R_i^2, 
\end{align}
with rotation operator $\R\!=\!\p\times\!\nabla_{\!\p}$
\cite{morse1953methods}, $\beta = 1/(k_{\rm B}T)$, and
an external force $\F_i$ which is absent in what follows. 
The active part of the dynamics is described by the operator $\delom_{\rm a} = -\sum_i \nab_{i}\!\cdot (v_0(\rr_i,t)\p_i)$. We refer to Refs.~\cite{sharma2016communication,sharma_PRE17} for a detailed elaboration of 
the linear response formalism. Here, we only give a brief outline of the method. We obtain from Eq.~\eqref{fp} an exact expression for 
the non-equilibrium average of a test function $f \equiv f(\rr^N\!\!,\p^N\!)$ as
\begin{align}\label{faverage}
\langle f \rangle(t) = \langle f \rangle_{\rm eq} - \int_{-\infty}^{t}\!\!dt'\,
\langle G(t') e_{-}^{\int_{t'}^{t}ds\,\omdag_{\rm a}(s)}f \rangle_{\rm eq}, 
\end{align}
where where $e_{-}$ is a negatively ordered exponential function~\cite{brader2012first}. We have defined $G(t) = K(t) + V(t)$ with
\begin{align}
K(t) &= \sum_{i=1}^{N} v_0(\rr_i,t)\,\p_i \cdot \beta\F_i \label{KP},\\
V(t) &= \sum_{i=1}^{N} \p_i\cdot \nabla_i v_0(\rr_i,t)  
\label{VP}
\end{align}
 and the adjoint operator is given by 
$\omdag_{\rm a}(t)=\omdag_{\rm eq}-\delom_{\rm a}(t)$, where $\omdag_{\rm eq}=\sum_{i} 
D_\text{t}\!\left(\nab_{i}\! + \beta\F_i\right) 
\!\cdot\!\nab_{i} \!+\! D_\text{r}\R_i^2$.   
Linear response corresponds to the system response when the full-time
evolution operator in~\eqref{faverage} is replaced by the time-independent
equilibrium adjoint operator. This is equivalent to assuming that the active
system is close to the equilibrium and the activity corresponds to a small  
perturbation.

\section{Polarization in the activity field}
\label{sec:orientation}
%In the following set of simulations, 500 active particles are
%placed into a box of the size $L^3 = 15^3$  and
%periodic boundaries. The particles have a diameter of $b=1$,
%but no interaction, so that actually an ensemble of single
%particles is simulated.
The activity field is a time-dependent
function of the $z-$ coordinate, which defines the magnitude
of the driving force as 
\begin{equation}\label{eq:f(z)}
f(z,t) = f_0 \left\{ \sin \left[\omega (z - v t)\right] + s\right\}\;,
\end{equation}
with the factor $f_0 = 2.5$, the system time $t$, the phase
velocity $v$ and the shift $s = 1.0$. A vertical shift of 
$s \ge 1$ is required to avoid unphysical negative values for 
the activity. The periodicity is accounted for with the choice
\begin{equation}
\omega = \frac{2n\pi}{L}
\end{equation}
and a positive integer $n$. The average orientation per particle
is defined as
\begin{equation}
\bm {p}(\bm{r}) = \frac{\langle \sum_i
\delta(\bm{r}-\bm{r}_i)\bm{p}_i\rangle}{\rho(\bm{r})}\;,
\end{equation}
with the one-body density $\rho(\bm{r}) =\langle \sum_i
\delta(\bm{r}-\bm{r}_i)\rangle$. The steady-state average orientation corresponding to a time-independent inhomogeneous activity field has been already obtained in Ref.~\cite{sharma_PRE17}. Corresponding to a space- and time-dependent activity, the steady-state orientation can be obtained using Eq.~\eqref{faverage} as

\begin{align}
\pp(\rr,t) &= \int_0^{t}\!dt'\int d\rr' v_0(\rr',t')\,\chi(|\rr - \rr'|,|t-t'|),
\label{pexpression}
\end{align}
where the space-time response function, $\chi(|\rr - \rr'|,t-t')$, 
%$\equiv \left(\frac{\delta \p(\rr)}{\delta v_0(\rr')}\right)_{\rm v_0 = 0}$ 
is given by
\begin{align}
\label{response}
\chi(|\rr - \rr'|,|t-t'|) =\frac{e^{-2D_{r}|(t-t')|}}{3}\nab G_{\rm VH}^{\rm s}(|\rr-\rr'|,|t-t'|).
\end{align}
In Eq.~\eqref{response}, $G_{\rm VH}(r,t)$ corresponds to the self-part of the Van Hove function. This function can be approximated as a Gaussian~\cite{Hansen90}
\begin{equation}
G_{\rm VH}^s(\bm{r},t) = \frac{1}{(4\pi D_t t)^{3/2}}\, e^{-r^2/4D_tt}.
\end{equation} 
We note that this approximation is valid even in the case of interacting
particles for any spherically-symmetric interaction potential, provided the
density is sufficiently low~\cite{sharma_PRE17}. Therefore, the presented
result is a generic one and not limited to ideal gases. In this work, we have
considered non-interacting particles as a special case, for which the Gaussian
approximation for the Van Hove function is exact.

\begin{figure}[t]
\includegraphics[angle=270,width=\columnwidth]{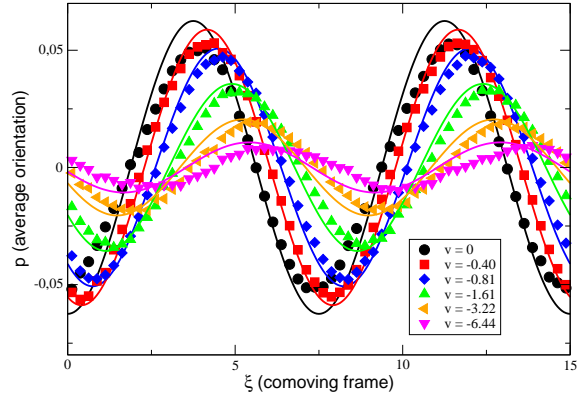}
\caption{Average orientation of the ABP in the coordinate 
system which moves with the phase velocity $v$ of the activity wave. Symbols
are MD simulations, closed curves are the approximation Eq.\  \eqref{eq:p}.
In this example, $n=2$ and $s=1$.}
\label{fig:orientation}
\end{figure}

Using Eqs.~\eqref{pexpression} and \eqref{response}, the average orientation corresponding to the activity wave in Eq.~\eqref{eq:f(z)} is obtained as 
\begin{eqnarray} \nonumber
p(z,t) =
&-&\frac{f_0\,\omega}{3\zeta}\int_{0}^{\infty} dt'\,
\int_{-\infty}^{\infty} dz'\, \cos \left[\omega (z' - v(t-t')) \right]\\
&\cdot&
\frac{\exp \left[-2D_r t' - \frac{(z-z')^2}{4D_t t'}\right]}
{\sqrt{4\pi D_t t'}}\;,
\end{eqnarray}

%\begin{equation}
%\bm{p}(\bm{r}) = -\int_{0}^{\infty} dt \,\frac{e^{-2D_rt}}{3\zeta} 
%\int_{-\infty}^{\infty} d\bm{r'}\, \nabla f(\bm{r'})\, G_{\rm VH}^s(
%|\bm{r}-\bm{r'}|,t)\;, 
%\end{equation}
%

%in which the self-part of the Van Hove function is approximated
%as a Gaussian~\cite{Hansen90}
%\begin{equation}
%G_{\rm VH}^s(\bm{r},t) = \frac{1}{(4\pi D_t t)^{3/2}}\, e^{-r^2/4D_tt}\;. 
%\end{equation} 
%In our quasi-1-dimensional setup, this transforms into
%\begin{eqnarray} \nonumber
%p(z,t) =
%&-&\frac{f_0\,\omega}{3\zeta}\int_{-\infty}^{\infty} dt'\,
%\int_{-\infty}^{\infty} dz'\, \cos \left[\omega (z' - v(t-t')) \right]\\
%&\cdot&
%\frac{\exp \left[-2D_r t' - \frac{(z-z')^2}{4D_t t'}\right]}
%{\sqrt{4\pi D_t t'}}\;,
%\end{eqnarray}
%
%
%
yielding
\begin{equation}\label{eq:p}
p(z,t) = -\frac{f_0\, \omega \cos \left[ \omega (z - vt) + 
\psi \right]}{3 \zeta \sqrt{(2D_r + D_t\omega^2)^2 + v^2}}\;,
\end{equation}
where we have introduced the phase shift
\begin{equation}\label{eq:psi}
\psi = \arctan\left[\frac{v}{2D_r + D_t \omega^2} \right]\;.
\end{equation}
The walker's orientation, Eq.\ \eqref{eq:p}, is stationary in the 
comoving coordinate frame $\xi = z-vt$ and is shown in Fig.\
\ref{fig:orientation}: With increasing phase velocity, the
amplitude of the orientation is decreasing, while the phase
shift increases. This is immediately obvious in Eq.\ \eqref{eq:p},
which exhibits the phase velocity in its denominator, and
in Eq.\ \eqref{eq:psi} in which the phase shift increases
linearly with $v$ in its leading order Taylor term. With
increasing phase velocity of the activity wave, the activity
changes rapidly so that the particle is eventually unable
to respond to changes of the external field. The crucial time scale is
the rotational relaxation time of $\tau_r = 1$, and 
when the activity wave runs from its minimum to its
next maximum during that time, the orientation profile
is essentially turning flat. This happens at phase velocities
of the order of $|v_{\rm max}| \approx \pi / 2$.
The orientation is generally 
pointing against the activity gradient, i.e.\ the ABP is 
turning toward the direction in which activity 
decreases~\cite{sharma_PRE17}.

The linear response theory generally overestimates the degree
of orientation, as well as the amount of phase shift. The 
deviation from the simulation results increases with the magnitude
of the driving force, or the pre-factor $f_0$ in Eq.\
\eqref{eq:f(z)}. On the other hand, we have observed almost
perfect agreement with the simulation data when $f_0$ was
set smaller than 0.1. It is clear that the validity of 
the linear response approximation remains restricted to 
the regime of low activities.

\section{Density distributions}\label{sec:densities}

\begin{figure}[t]
		\includegraphics[angle=270,width=\columnwidth]{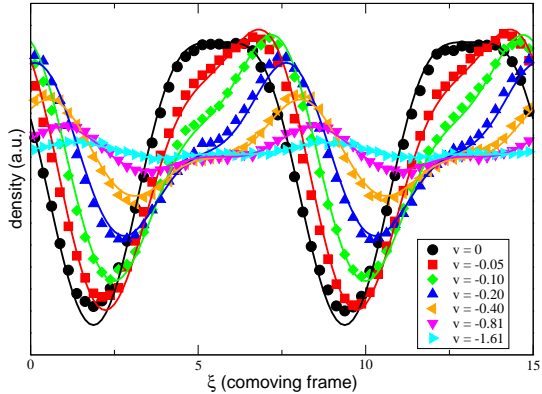}
		\caption{Density distributions of the ABPs in the coordinate 
comoving frame at different phase velocities of the activity wave. Symbols
are MD simulations, closed curves are the approximation Eq.\  \eqref{eq:rho1}.
In this example, $n=2$ and $s=1$.}
\label{fig:density}
\end{figure}

Traveling activity waves as in Eq.~\eqref{eq:f(z)} induces traveling
orientation waves (Eq.~\eqref{eq:p}) and traveling density waves. One cannot,
however, use linear response approach to calculate the density distribution
because it is invariant to activity in the linear order. To calculate density
distribution, we integrate over $\p$, $x$ and $y$ in Eq.~\eqref{fp} to obtain
the following 1-dimensional equation for the density $\rho(z,t)$
\begin{equation}\
\frac{\partial \rho(z,t)}{\partial t} = \frac{\partial}{\partial z}\left[D_{t}\frac{\partial \rho(z,t)}{\partial z} - \frac{f_0(z,t)}{\zeta}p(z,t) \rho(z,t) \right].
\label{eq:density}
\end{equation}
In the comoving frame $\xi$, Eq.~\eqref{eq:density} can be recast as a continuity equation
\begin{equation}\label{eq:delrho} 
\frac{\partial \rho}{\partial t} + \frac{\partial J}{\partial \xi}
= 0\;,
\end{equation}
where $\rho = \rho(z-vt) \equiv \rho(\xi)$ and $J(\xi)$ is the flux in the comoving frame given as
\begin{equation}\label{Jflux}
J(\xi) = -D_t \frac{\partial \rho}{\partial \xi} + \frac{f(\xi)}{\zeta}\, p(\xi)\, \rho(\xi)
- v\, \rho(\xi).
\end{equation}
Since the density is stationary in the $\xi$-frame, it follows from Eq.~\eqref{eq:delrho} that  $J(\xi)$ is constant. On integrating Eq.~\eqref{Jflux} and using periodicity of $p(\xi)$ and $\rho(\xi)$, one obtains the following equations for the flux
\begin{equation}\label{eq:j2}
\frac{J}{\rho_b L D_t} = \frac{\left[
1 - \exp\left\{
-\int_0^L \frac{b(z)}{D_t} \, dz
\right\}
\right]
}{
\int_0^L dx\, \int_0^L dy\, \exp \left\{
-\int_x^{x+y} \frac{b(z)}{D_t} \, dz
\right\}
}\;
\end{equation}
and the density
\begin{equation}\label{eq:rho1}
\frac{\rho(\xi)}{\rho_b} = \frac{L
\int_0^L dy\, \exp\left[ 
- \int_{\xi}^{\xi+y} \frac{b(\xi')}{D_t}\, d\xi'
\right]
}{
\int_0^L d\xi\, \int_0^L dy\, \exp \left[
- \int_{\xi}^{\xi+y} \frac{b(\xi')}{D_t}\, d\xi'
\right],
}
\end{equation}
with the particle bulk density $\rho_b = N/V$ and the function
$b(\xi) = \zeta^{-1} f(\xi) \, p(\xi) - v$.

%In linear response, the density distribution of the ABPs
%in the comoving frame ($\xi = z-vt$) is expressed as
%\begin{equation}\label{eq:rho1}
%\frac{\rho(\xi)}{\rho_b L} = \frac{
%\int_0^L dy\, \exp\left[ 
%- \int_{\xi}^{\xi+y} \frac{b(\xi')}{D_t}\, d\xi'
%\right]
%}{
%\int_0^L d\xi\, \int_0^L dy\, \exp \left[
%- \int_{\xi}^{\xi+y} \frac{b(\xi')}{D_t}\, d\xi'
%\right]
%}
%\end{equation}
%with the particle bulk density $\rho_b = N/V$ and the function
%\begin{equation}
%b(\xi) = \zeta^{-1} f(\xi) \, p(\xi) - v\;.
%\end{equation}
%\begin{equation}
%b(\xi) = \frac{-\beta}{2}\left[
%\sin(2\omega\xi+\psi) - \sin \psi + 2 s \cos(\omega\xi+\psi)
%\right] - v
%\end{equation}
%with the factor
%\begin{equation}
%\beta = \frac{f_0^2 \omega}{3\zeta^2} \left[
%(2D_r + D_t \omega^2)^2 + \omega^2 v^2
%\right]^{-1/2}\;.
%\end{equation}
This integral (Eq.~\eqref{eq:rho1}) has to be solved numerically, solutions are 
shown in Fig.\ \ref{fig:density}. We have used the theoretical prediction of Eq.~\eqref{eq:p} for $p(\xi)$. The linear response theory
offers an excellent approximation to the simulation data in
the present parameter ranges. Even at high phase
velocities, the density distributions are almost
quantitatively reproduced. A comparison to  Fig.\ 
\ref{fig:orientation} reveals the different time scales at which
the variables respond to the activity field: While the density
distribution is almost uniform at phase velocities as low as
$|v| \approx 1.6$, the polarization pattern is not yet flat
at a far higher velocity of $|v| \approx 6.4$. 
This is a consequence of the longer relaxation times
associated with translocations of the ABPs, as
compared to their rotational relaxation times. In the
present parameter setting, the particle positions relax
roughly an order of magnitude slower than their orientations.    

\section{Induced flux}\label{sec:flux}

\begin{figure}[t]
		\includegraphics[angle=270,width=\columnwidth]{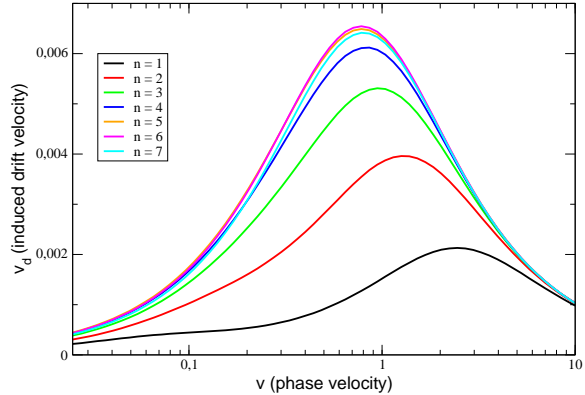}
		\caption{Induced drift velocities
                  (Eq.\ \ref{eq:vd}), as a function of the phase velocity of
the activity wave. Different curves correspond to different angular
frequencies $\omega = 2n\pi/L$ that satisfy the periodic boundary condition.
Here, the size of the ABP equals $b=1$, the activity factor $f_0 = 2.5$.
}
\label{fig:omega_b1}
\end{figure}

%While propagating through the system, the activity wave induces
%a flux into the ensemble of ABPs. In the comoving coordinate frame,
%the dynamic equation for the density distribution is
%\begin{equation}\label{eq:delrho} 
%\frac{\partial \rho}{\partial t} = \frac{\partial}{\partial \xi}
%\left[
%D_t \frac{\partial \rho}{\partial \xi} - f(\xi)\, p(\xi)\, \rho(\xi)
%+ v\, \rho(\xi)
%\right] = 0\;,
%\end{equation} 
%since the density is stationary in comoving coordinates. This is leading
%to the flux 
%\begin{equation}\label{eq:j1}
%-J = D_t \frac{\partial \rho}{\partial \xi} - b(\xi)\, \rho(\xi)\;,
%\end{equation}
%which is partially integrated to yield
%\begin{equation}\label{eq:j2}
%J = \frac{\rho_b L D_t \left[
%1 - \exp\left\{
%-\int_0^L \frac{b(z)}{D_t} \, dz
%\right\}
%\right]
%}{
%\int_0^L dx\, \int_0^L dy\, \exp \left\{
%-\int_x^{x+y} \frac{b(z)}{D_t} \, dz
%\right\}
%}\;,
%\end{equation}
The average drift velocity
of the ABPs in laboratory frame can be written as
\begin{equation}\label{eq:vd}
v_d = \frac{J}{\rho_b} + v\;.
\end{equation}

Figure \ref{fig:omega_b1} displays solutions to Eq.\ \ref{eq:vd} for 
different angular frequencies $\omega = 2n\pi/L$ of the activity waves.
As a function of the phase velocity, the induced drift generally 
exhibits a global maximum, which, however, differs with the angular
frequency. The overall maximum value of the drift velocity is found
with $n = 6$ and a phase velocity about $v \approx 0.75$. We note that 
an earlier study has reported on situations in which negative
drifts, i.e.\ in the direction opposite to the propagation of the
activity wave~\cite{geiseler_PRE16,geiseler_SR17}. So far, we have 
not been able to reproduce such a phenomenon with our setup in 
three dimensional systems, neither in linear response approximation
nor in simulation.  It may well be the case that the parameter range
covered in our studies did not allow for such a reversal of the 
direction of drift.

\begin{figure}[t]
		\includegraphics[angle=270,width=\columnwidth]{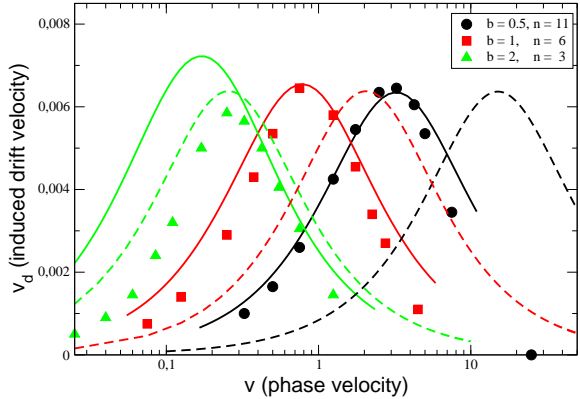}
		\caption{Induced drift velocities as a function of phase
                  velocity
for different particle sizes and angular frequencies $\omega = 2n\pi/L$. Data
points are simulations, solid curves computed from Eq.\ \eqref{eq:vd},
and dashed curves are the approximation \eqref{eq:vd_analyt}.}
		\label{fig:vd_vs_vp_all}
\end{figure}

It is natural to ask how an induced drift would depend upon
the properties of the particle. We have repeated the analysis of
 Fig.\ \ref{fig:omega_b1} with particles of sizes $b=0.5$ and
$b=2$, with properties as summarized in Tab.\ \ref{tab:parameters}.
Figure \ref{fig:vd_vs_vp_all} only contains curves for angular frequencies
at which the respective particles reach the highest drift velocities.
For small ABPs, MD simulations and linear response theory (Eq.\ \eqref{eq:vd})
display a reasonably close agreement. For the larger particle of 
diameter $b=2$ (green curve and triangles), however, significant
deviations are visible. Due to their longer rotational relaxation
times, the dynamics of these particles are less
affected by Brownian motion, i.e.\ more ballistic, and linear 
response theory begins to break down. In Sec.\ \ref{sec:activity}
we are going to discuss the issue of valididy ranges
of the linear response approach in detail.

It is instructive to investigate an approximation to the induced
drift velocity, considering that the orientation of the particle
relaxes faster than its density profile. If the traveling wave 
propagates sufficiently fast, then $\rho(\xi)$ in Eq.~\eqref{Jflux}
may be approximated by the uniform bulk density, turning 
Eq.~\eqref{eq:vd} into
\begin{equation}\label{eq:vd_app}
v_d(\xi) \approx \frac{1}{\zeta} p(\xi)\, f(\xi)\;, 
\end{equation}
which can be integrated over the box to yield the average drift velocity
\begin{equation}\label{eq:vd_analyt}
v_d = \frac{f_o^2 \omega v}{6 \zeta^2 \left[(2 D_r + D_t \omega^2)^2 + v^2 \right]}\;.
\end{equation}
The plots in Fig.\ \ref{fig:vd_vs_vp_all} show that this approximation
(dashed curves) is reasonable only in case of the large particles, which
display a sufficiently slow reaction to the incoming activity field so
that $\rho(\xi) \approx \rho_b$ remains valid. Yet, Eq.~\eqref{eq:vd_analyt}
provides us with a glimpse of how the underlying dynamics of the ABPs 
enables the induction of drift: This function has its extremum at
\begin{equation}\label{eq:maxima}
\tilde{v} = 4D_r \sim b^{-3}\;,\;\;\;\; \tilde{\omega} = 
\sqrt{\frac{2D_r}{D_t}}\sim b^{-1}\;,
\end{equation}
at which it reaches the induced drift velocity of
\begin{equation}
v_{d, \rm max} = \frac{\sqrt{2}}{48}\, 
\frac{f_o^2}{\zeta^2 \sqrt{D_r D_t}} \sim b^0\;,
\end{equation}
where we used the fact that $\zeta \sim b$. The optimum choice for the velocity
of the traveling wave is determined by the rotational 
diffusion and thus strongly dependent on the particle size. While
the $b^{-3}$-scaling of this approximation to the optimum
drift velocity definitely overestimates the simulation results,
for which the scaling is closer to $b^{-2}$, the ideal
angular frequency is inversely proportional to $b$ in
both, approximation and simulation. The
peak drift velocity is not an explicit function of the particle
size, which is as well supported by the simulations 
(Fig.\ \ref{fig:vd_vs_vp_all}) and thus no spurious consequence of
the uniform-density-approximation. 
In the laboratory, of course, the driving force $f_o$ is likely
to depend on the particle diameter and in this way the 
maximum achievable drift 
velocities may turn species-dependent.

\section{Separation of mixtures of ABPs with different sizes}
\label{sec:separation}

\begin{figure}[t]
\includegraphics[width=\columnwidth]{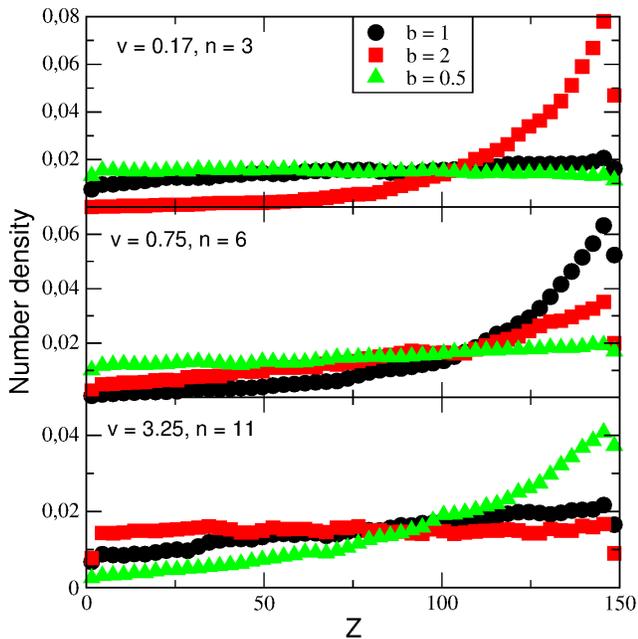}
\caption{A mixture of three ABP-species of different
  diameters. Depending on the choices for the phase velocity $v$ and the angular 
  frequency $\omega = 2n\pi/L$ of the activity wave, either the large
particles (upper panel), the medium sized particles (center panel) or
the small particles (lower panel) are enriched near the wall.} 
\label{fig:threesizes_b1}
\end{figure}

In this set of simulations, the box-length was increased to $150$ 
in z-direction, where the boundaries were fixed with short-range 
repulsive walls. Three particle species were added, 500 each,
with parameters as summarized in Tab.\ \ref{tab:parameters}. No 
interactions between particles were enabled, so that these are 
pure averages of single particle ensembles. Then, activity waves 
were sent through the system, with positive phase velocities
(from the left to the right in Fig. \ref{fig:threesizes_b1})
parameters at which the solid curves of 
Fig.\ \ref{fig:vd_vs_vp_all}, e.g.\ Eq.~\eqref{eq:vd}, had 
their maxima.

As can be seen in Fig.\ \ref{fig:vd_vs_vp_all}, the different values
of induced drift velocities allow for a separation of the mixture
of particle species: Either one of the three species may be enriched
at the right hand wall, depending on the particular choice of the
activity wave. It is therefore possible to separate a mixture of 
ABPs according to their diameter. In the present simulation, the
density distributions of the enriched species had turned stationary 
after simulation times of roughly $10^4$, so that a continuation
of that procedure did not improve its selectivity any further.

\section{Validity of the linear response approach}\label{sec:activity}
The solution to Eq.~\eqref{fp} is based on a separation
of the time evolution operator $\om_{\rm a}(t)=\omeq+\delom_{\rm a}(t)$
into its (diffusive) equilibrium part and a second, activity
driven part, which is treated as a perturbation. This linear
response ansatz produces accurate dynamics as long as diffusion 
dominates over driven motion. 

To compare both components, we consider the persistence length
of ballistic motion, $l_b = v_b \tau_r = f\tau_r/\zeta$, with
the average diffusion distance $l_d = \sqrt{6D_t\tau_r}$ during
the same time interval $\tau_r$. As long as $l_b/l_d \ll 1$,
diffusion dominates the motion. In our present simulations,
the peak driving force reaches $f = 5$, at which
 $l_b/l_d = 1/4$ ($b=1/2$), $l_b/l_d = 1/2$ ($b=1$) and
$l_b/l_d = 1$ ($b=2$), which indicates that the formalism
is likely to break down in case of the large particles,
as is obvious in Fig.\ \ref{fig:vd_vs_vp_all}.

A diffusion driven dynamics is not uncommon for ABPs of 
the sub - $100$ $nm$ scale. As an example, we consider the 
dynamics of Au-Pt Janus particles of diameter $30$ nm, as prepared
and analyzed by Lee et al.~\cite{lee_NanoLett14}: In a solution
of water with 2.5\% H$_2$O$_2$, these particles developed
a considerable ballistic speed of $2.2\cdot 10^4$ particle 
diameters per second. However, due to their small diameters,
the rotational diffusion was rapid and allowed for a
persistent motion over a distance of only $l_b \approx 4.6$ $nm$
during the directional persistence time of 
$\tau_r \approx 7$ $\mu s$. Given the diffusion coefficient 
of the passive particle, $D_t \approx 0.013$ $nm^2\cdot ns^{-1}$,
the distance of $l_d = \sqrt{6D_t \tau_r} \approx 23$ $nm$ 
was covered during one persistence time. This yields the ratio
$l_b/l_d \approx 1/5$, so that linear response theory may
safely be applied to this system.  
 
We summarize that the linear response approximation
describes ABPs at activity levels that are comparable or
smaller than the diffusive motion of the particle
and may hence be a method of choice for small ABPs
of diameters in the sub - $0.1$ $\mu m$ range, as well as 
for larger particles which are only weakly driven.

\section{Summary}\label{sec:summary}

In the present work, a linear response (Green-Kubo)
approach has been applied to ABPs
that are exposed to time-varying activity fields. We
have demonstrated how the distribution of particle orientations, 
their densities and the resulting induced fluxes are approximated,
and shown that these approximations agree closely with Langevin
dynamics simulations (Sections \ref{sec:orientation} -- 
\ref{sec:flux}).

The accuracy of linear response theory is granted as long as
the contribution from their active motion remains weak, so that 
their diffusive thermal motion is dominating the dynamics. This 
is naturally the case with small ABPs of sub- 100 $nm$ diameter, 
or with larger particles that are only weakly driven (Sec.\ \ref{sec:activity}).

Dynamic activity waves are capable of inducing fluxes into systems
of ABPs, and the efficiency of that coupling is a function of
the particle diameter, as is most easily seen in the 
approximation~\eqref{eq:maxima}. A tuning of the phase- and
angular velocity allows the practitioner to efficiently drive
particles of selected diameters through the system, which might
be used for a controlled separation
of mixtures of ABPs according to their sizes (Sec.\ \ref{sec:separation}). 

Within the accuracy of the formalism, linear response theory yields
the parameters for the activity waves which may be applied to facilitate 
such a directed transport at highest efficiency.

%\section*{Acknowledgements}

%\bibliographystyle{unsrt}
%\bibliography{active}

\end{document}